\documentclass[twocolumn,superscriptaddress]{revtex4-1}

\usepackage[dvipdfmx]{graphicx}
\usepackage{dcolumn}%
\usepackage{bm}%
\usepackage{multirow}

\usepackage{color}
\usepackage{ascmac}

\begin{document}

\title{
Efficient algorithm based on Liechtenstein method for computing exchange coupling constants using localized basis set 
}
\author{Asako Terasawa}
\email{terasawa.a.aa@m.titech.ac.jp}
\affiliation{
Department of Materials Science and Engineering, Tokyo Institute of Technology,
J1-3, Nagatsuta-cho 4259, Midori-ku, Yokohama 226-8502, Japan}
\author{Munehisa Matsumoto}
\affiliation{
Institute for Solid State Physics, The University of Tokyo,
Kashiwanoha 5-1-5, Kashiwa 277-8581, Japan}
\author{Taisuke Ozaki}
\affiliation{
Institute for Solid State Physics, The University of Tokyo,
Kashiwanoha 5-1-5, Kashiwa 277-8581, Japan}
\author{Yoshihiro Gohda}
\affiliation{
Department of Materials Science and Engineering, Tokyo Institute of Technology,
J1-3, Nagatsuta-cho 4259, Midori-ku, Yokohama 226-8502, Japan}

\begin{abstract}
For large-scale computation of the exchange coupling constants $J_{ij}$, we reconstruct the Liechtenstein formula for localized orbital representation and simplify the energy integrations by adopting the finite pole approximation of the Fermi function proposed by Ozaki [Phys. Rev. B \textbf{75}, 035123 (2007)].
We calculate the exchange coupling constant $J_{\mathrm{1NN}}$ of the first-nearest-neighbor sites in body-centered-cubic Fe systems of various sizes to estimate the optimal computational parameters that yield appropriate values at the lowest computational cost.
It is shown that the number of poles needed for a computational accuracy of 0.05 meV is determined as $\sim$ 60, whereas the number of necessary Matsubara poles needed to obtain similar accuracy, which was determined in previous studies, is on the order of 1000.
Finally, we show $J_{ij}$ as a function of atomic distance, and compared it with one derived from Korringa-Kohn-Rostoker Green's function formalism. The distance profile of $J_{ij}$ derived by KKR formalism agrees well with that derived by our study, and this agreement supports the reliability of our newly derived formalism.
\end{abstract}

\maketitle

\section{Introduction}\label{sec:intro}
Considering recent industrial demands, detailed study of the structural and electronic properties of magnetic materials has become an important issue in materials science.
In particular, theoretical understanding of permanent magnets is one of the most important yet difficult topics, because of the fundamental complexity of permanent magnets.
Recent studies of permanent magnets have revealed complicated material microstructure in Nd--Fe--B-type permanent magnets \cite{S_Sugimoto_2011, K_Hono_2012,S_Hirosawa_2017}.
Specifically, grain boundaries are crucial to enhancing the coercivity of permanent magnets because of their pinning effect, which prevents the movement of magnetic domain walls \cite{S_Sugimoto_2011, K_Hono_2012, S_Hirosawa_2017, S_Li_2002, W_F_Li_2009, T_H_Kim_2012, H_Sepehri_Amin_2012, U_M_R_Seelam_2016}.
Moreover, it was reported that both the crystallinity and composition ratio of the grain boundary phase change depending on the relative angles with respect to the $c$ axes of neighboring grains of the main phase \cite{T_T_Sasaki_2016}. 

Recently, first principles computation techniques have been employed to investigate the magnetic properties of permanent magnets \cite{B_Balasubramanian_2016, A_Saengdeejing_2016, Y_Tatetsu_2016, Z_Torbatian_2016, N_Umetsu_2016, H_Akai_2018, Y_Gohda_2018, C_E_Patrick_2018, Y_Tatetsu_2018, C_E_Patrick_2019, A_M_Schonhobel_2019, A_L_Tedstone_2019}. It is, however, still difficult to understand the details of magnetic interaction in the microstructures of permanent magnets because of their structural complexity.


The Liechtenstein method is a powerful tool that uses second-order perturbation theory to treat the magnetism of complicated systems by examining the response of the total energy to rotation of the spins at two atomic sites through infinitesimal angles \cite{A_I_Liechtenstein_1987}. The original Liechtenstein formula is written as follows:
\begin{equation}
J_{ij}
=\frac{1}{4\pi}\int d\varepsilon f\left(\beta(\varepsilon-\varepsilon_{\mathrm{F}})\right)
\mathrm{Im\,Tr}\left[
\hat{\Delta}_{i}\hat{T}^{ij}_{\uparrow}\hat{\Delta}_{j}\hat{T}^{ji}_{\downarrow}
\right],
\label{eq:Liech_path}
\end{equation}
where $T^{ij}_{\sigma}$ is the path integral operator between sites $i$ and $j$, $\hat{\Delta}_{i}\equiv\hat{t}_{i\uparrow}-\hat{t}_{i\downarrow}$ is the single-site scattering matrix $\hat{t}_{i\sigma}$ at site $i$, $f(x)=1/[\exp(x)+1]$ is the Fermi function with $\beta = (k_{\mathrm{B}}T)^{-1}$, and $\varepsilon_{\mathrm{F}}$ is the Fermi level.
This formula was also rewritten within the Green's function formalism as follows \cite{M_Pajda_2001,I_Turek_2003}:
\begin{eqnarray}
J_{ij}
&=
 &\frac{1}{4\pi}\int d\varepsilon\,f\left(\beta(\varepsilon-\varepsilon_{\mathrm{F}})\right)\,
\nonumber \\
&&\quad\times\mathrm{ImTr}\left[
    \hat{G}_{\uparrow}^{+}(\varepsilon)\hat{P}_{i}
    \hat{G}_{\downarrow}^{+}(\varepsilon)\hat{P}_{j}
  \right],
\label{eq:Liech_Green}
\end{eqnarray}
\begin{equation}
\hat{P}_{i} \equiv \hat{H}_{i\uparrow} -\hat{H}_{i\downarrow},
\label{eq:pot_diff}
\end{equation}
where $\hat{G}_{\sigma}^{+}(\varepsilon)$ is the retarded Green's function of the spin $\sigma$ in unperturbed states, and $\hat{H}_{i\sigma}$ is the on-site term of the Hamiltonian for spin $\sigma$ at site $i$.

Owing to use of the Liechtenstein formulae and the development of first-principles calculations of magnetic materials, the magnetic properties of various magnetic materials have begun to be revealed  \cite{V_A_Dinh_2009,I_Galanakis_2011,M_Seike_2012,T_Fukushima_2015}.
It is, however, difficult to understand the magnetic properties of various phases and their interfaces in permanent magnets because the structural details of subphases in permanent magnets are still under investigation  \cite{T_T_Sasaki_2016,X_D_Xu_2018}, and the interfacial structures between multiple phases tend to become very complicated for large numbers of atoms.
Therefore, quantitative understanding of the magnetism of an entire permanent magnet contains many related problems to be solved, and the formulation of the exchange coupling constants for large-scale computation is a significant problem associated with this challenging task.

In this paper, we derive the explicit forms of the Liechtenstein formula on the basis of non-orthogonal localized orbitals for the interaction between two individual sites located in different cells and for the interaction between periodic images. For the former derivation, we simplify the formalism by adopting the approximate form of the Fermi function proposed by Ozaki \cite{T_Ozaki_2007}.
We implement this formalism in an MPI code that uses the overlap and Hamiltonian matrices of OpenMX \cite{T_Ozaki_2003}, a first-principles calculation code based on a linear combination of pseudoatomic orbitals approximation.
To determine the most efficient computational conditions, we examine the calculated values of $J_{\mathrm{1NN}}$ of the first-nearest-neighbor (1NN) sites in body-centered cubic (bcc) Fe crystals in systems of various sizes and under different computational conditions.
We find that it is necessary to take approximately 60 poles of the approximated Fermi functions, whereas the number of Matsubara poles needed for sufficient accuracy was reported to be approximately 1000 or more by Kvashnin \textit{et al.} \cite{Y_O_Kvashnin_2015}.
Finally, we calculate the dependence of $J_{ij}$ on the atomic distance $r_{ij}$, and compared it with a $J_{ij}$ profile of bcc Fe obtained by Akai-KKR, a first-principles calculation code based on Korringa-Kohn-Rostoker (KKR) Green's function formalism \cite{H_Shiba_1971,H_Akai_1977H_Akai_1982,J_Korringa_1947,W_Kohn_1954}.
It is shown that the $J_{ij}$ profile in this study agrees well with that derived by Akai-KKR. Considering the variation of calculated $J_{ij}$ among different previous studies \cite{M_Pajda_2001,H_Hang_2010,Y_O_Kvashnin_2015,H_Yoon_2018}, it is possible to say that the agreement between two formalism support the reliability of calculated values with each other.
\section{Modification of Liechtenstein formula}\label{sec:Liech}

\subsection{Non-orthogonal basis representation}

\begin{figure}[hbtp]
\begin{center}
\includegraphics[width=80mm]{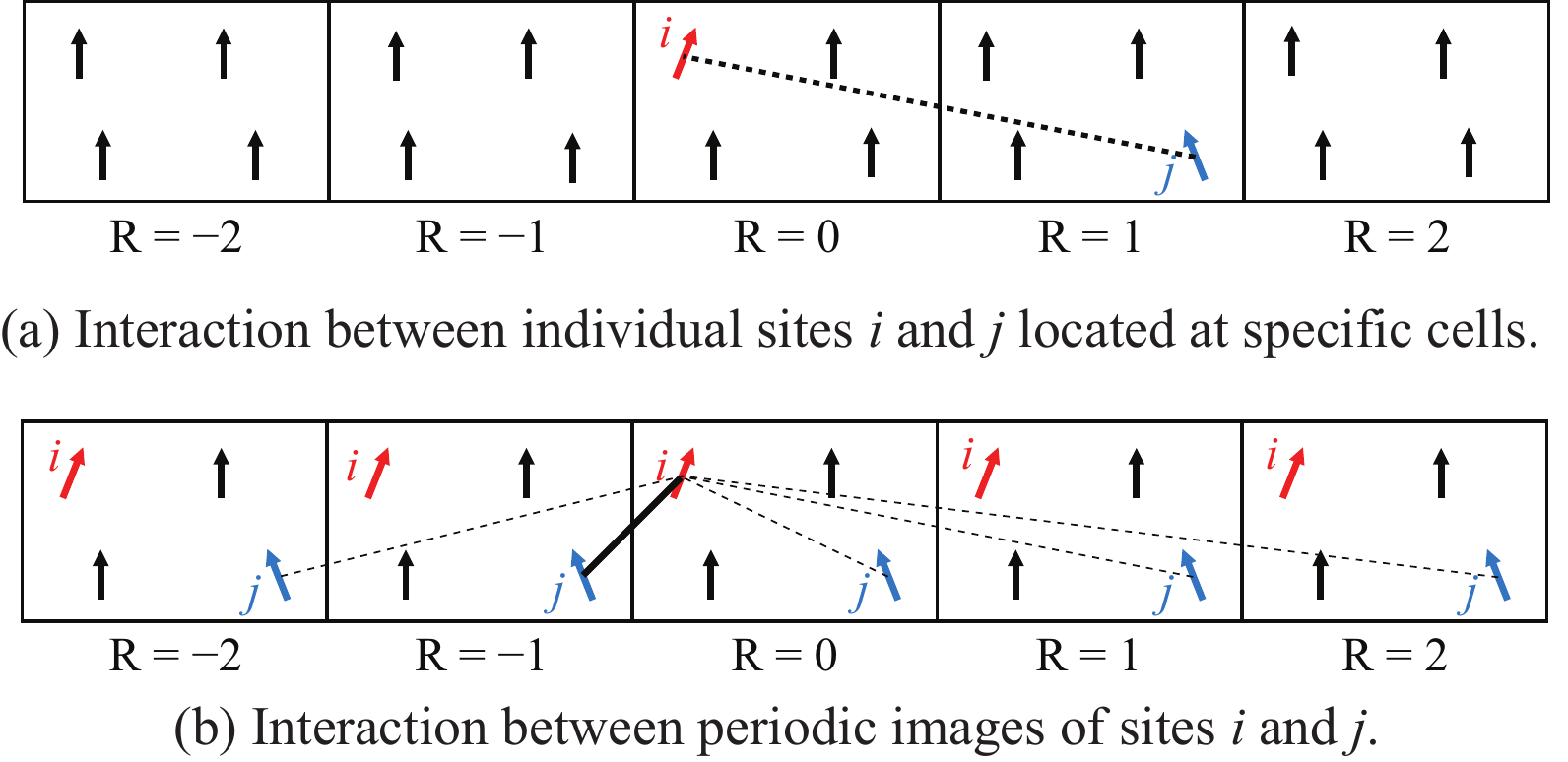}\\
\end{center}
\caption{Schematic of interaction between periodic images of sites $i$ and $j$, and that between individual sites $i$ and $j$ located in specific cells. 
\label{fig:sch_cells}
}
\end{figure}

Our aim in this section is to represent the Liechtenstein formula by a localized basis set with an efficient algorithm for large-scale computations.
As a starting point, let us consider the exchange coupling constant between site $i$ in cell $\mathbf{0}$ and site $j$ in cell $\mathbf{R}$, which are shown in Fig.~\ref{fig:sch_cells}(a). With the explicit cell indices $\mathbf{0}$ and $\mathbf{R}$, Eqs.~(\ref{eq:Liech_Green}) and (\ref{eq:pot_diff}) can be rewritten as 
\begin{eqnarray}
J_{i\mathbf{0},j\mathbf{R}}
&=
 &\frac{1}{4\pi}\int d\varepsilon\,f\left(\beta(\varepsilon-\varepsilon_{\mathrm{F}})\right)
\nonumber \\
&&\quad \times
  \mathrm{ImTr}\left[
    \hat{G}_{\uparrow}^{+}(\varepsilon)\hat{P}_{i\mathbf{0}}
    \hat{G}_{\downarrow}^{+}(\varepsilon)\hat{P}_{j\mathbf{R}}
  \right],
\label{eq:Jij_indiv}
\end{eqnarray}
\begin{equation}
\hat{P}_{i\mathbf{R}}
\equiv
\hat{H}^{(\mathrm{KS})}_{i\mathbf{R}}(\uparrow)
-\hat{H}^{(\mathrm{KS})}_{i\mathbf{R}}(\downarrow),
\end{equation}
where $\hat{H}^{(\mathrm{KS})}_{i\mathbf{R}}(\sigma)$ represents the on-site partial matrix of the Kohn--Sham Hamiltonian of spin $\sigma$ at site $i$ of cell $\mathbf{R}$.
To treat Eq.~(\ref{eq:Jij_indiv}) with a localized orbital basis set, we used the expansion of Bloch functions by the localized orbitals,
\begin{equation}
|\mathbf{k},n,\sigma\rangle
=\frac{1}{\sqrt{N}}\sum_{\mathbf{R}}\sum_{i}\sum_{\mu\in i}
|\mathbf{R},i,\mu\rangle e^{i\mathbf{k}\cdot\mathbf{R}} C_{i\mu,n\sigma}(\mathbf{k}),
\label{eq:PAO}
\end{equation}
where $\mu$ represents the index of the orbital belonging to site $i$, and $N$ represents the number of cells related to the periodic boundary condition.
As a result, Eq.~(\ref{eq:Jij_indiv}) can be written in the following explicit form:
\begin{eqnarray}
J_{i\mathbf{0},j\mathbf{R}}
&=
 &\frac{1}{4\pi N^2}\sum_{\mathbf{k},\mathbf{k}'}
  \sum_{n,n'}
  \sum_{\mu,\nu\in i}\sum_{\mu',\nu' \in j}
  e^{i(\mathbf{k}-\mathbf{k}')\cdot\mathbf{R}}
\nonumber \\
&&\times
  \int d\varepsilon\,f\left(\beta(\varepsilon-\varepsilon_{\mathrm{F}})\right)
\nonumber \\
&&\quad\times\,\mathrm{Im}\left[
  \frac{C_{j\mu',n\uparrow}(\mathbf{k})C^{*}_{i\nu,n\uparrow}(\mathbf{k})}{\varepsilon+i\eta-\varepsilon_{n\uparrow}(\mathbf{k})}
  [\hat{P}_{i}]_{\nu\mu}\right.
\nonumber \\
&&\left. \qquad \times
  \frac{C_{i\mu,n'\downarrow}(\mathbf{k}')C^{*}_{j\nu',n'\downarrow}(\mathbf{k}')}
       {\varepsilon+i\eta-\varepsilon_{n'\downarrow}(\mathbf{k}')}
  [\hat{P}_{j}]_{\nu'\mu'}
  \right],
\label{eq:Jij_indiv_loc}
\end{eqnarray}
where $\varepsilon_{n\sigma}(\mathbf{k}),\mathbf{C}_{n\sigma}(\mathbf{k})$ 
represent the corresponding eigenvalue and eigenvector, respectively, indexed by $n$ and $\sigma$ for the Kohn--Sham equation at the wave number $\mathbf{k}$; and $[\hat{P}_{i}]_{\nu\mu}$ and $[\hat{P}_{j}]_{\nu'\mu'}$ represent the partial matrices of the potential difference operator at sites $i$ and $j$, respectively.
We show that Eq.~(\ref{eq:Jij_indiv_loc}) is relevant not only for orthogonal basis sets but also for non-orthogonal basis sets; a detailed derivation is given in Appendix \ref{append:no_Liech}.

There are two ways to simplify Eq.~(\ref{eq:Jij_indiv_loc}). One is the eigenfunction representation,
\begin{eqnarray}
J_{i\mathbf{0},j\mathbf{R}}
&=
 &\frac{1}{4}\int d^3\left(\frac{ka}{2\pi}\right)\int d^3\left(\frac{k'a}{2\pi}\right)e^{i(\mathbf{k}-\mathbf{k}')\cdot\mathbf{R}}
\nonumber \\
&&\quad\times
\sum_{n,n'}
  \frac{-f_{n\uparrow}(\mathbf{k})+f_{n'\downarrow}(\mathbf{k}')}
       {\varepsilon_{n\uparrow}(\mathbf{k})-\varepsilon_{n'\downarrow}(\mathbf{k}')}
\nonumber \\
&&\quad\times
  \sum_{\mu,\nu\in i}\sum_{\mu',\nu' \in j}
   C_{j\mu',n\uparrow}(\mathbf{k})C^{*}_{i\nu,n\uparrow}(\mathbf{k})
  [\hat{P}_{i}]_{\nu\mu}
\nonumber \\
&&\qquad\qquad\times
  C_{i\mu,n'\downarrow}(\mathbf{k}')C^{*}_{j\nu',n'\downarrow}(\mathbf{k}')
  [\hat{P}_{j}]_{\nu'\mu'},
\label{eq:Jij_eig}
\end{eqnarray}
which can be obtained by replacing the imaginary parts of the Green's functions with delta functions around the energy eigenvalues \cite{M_J_Han_2004}.
Direct implementation of Eq.~(\ref{eq:Jij_eig}), however, may be computationally costly because of the double integrals of two wave vectors. 

The other representation of Eq.~(\ref{eq:Jij_indiv_loc}) is the Green's function representation, which is written as follows:
\begin{eqnarray}
J_{i\mathbf{0},j\mathbf{R}}
&=
 &\frac{1}{4\pi}
  \sum_{\mu,\nu\in i}\sum_{\mu',\nu' \in j} 
  \int_{-\infty}^{\infty} d\varepsilon\,f\left(\beta(\varepsilon-\varepsilon_{\mathrm{F}})\right)
\nonumber \\
&&\quad \times\,\mathrm{Im}\left\{
  [\hat{P}_{i}]_{\nu\mu}
  G^{+}_{i\mu,j\nu'}(\downarrow,\varepsilon,\mathbf{R})
  \right.
\nonumber \\
&&\qquad\qquad \times\left.
  [\hat{P}_{j}]_{\nu'\mu'}
  G^{+}_{j\mu',i\nu}(\uparrow,\varepsilon,-\mathbf{R})
  \right\},
\label{eq:Jij_Green}
\end{eqnarray}
where
\begin{eqnarray}
&&G^{+}_{j\mu',i\nu}(\uparrow,\varepsilon,-\mathbf{R})
\nonumber \\
&&=\int d^3\left(\frac{ka}{2\pi}\right)e^{i\mathbf{k}\cdot\mathbf{R}}\sum_{n}
  \frac{C_{j\mu',n\uparrow}(\mathbf{k})C_{i\mu,n\uparrow}(\mathbf{k})}
  {\varepsilon+i\eta-\varepsilon_{n\uparrow}(\mathbf{k})}
  \label{eq:g_up_2}
\\
&&G^{+}_{i\mu,j\nu'}(\downarrow,\varepsilon,\mathbf{R})
\nonumber \\
&&=\int d^3\left(\frac{ka}{2\pi}\right)e^{-i\mathbf{k}\cdot\mathbf{R}}\sum_{n'}
  \frac{C_{i\mu,n'\downarrow}(\mathbf{k})C_{j\nu',n'\downarrow}(\mathbf{k})}
  {\varepsilon+i\eta-\varepsilon_{n'\downarrow}(\mathbf{k})}.
\label{eq:g_down_2}
\end{eqnarray}
Equation (\ref{eq:Jij_Green}) involves the integral for the energy $\varepsilon$, and Eqs. (\ref{eq:g_up_2}) and (\ref{eq:g_down_2}) involve single integrals for the wave vector $\mathbf{k}$.

\subsection{Contour integrals and finite pole approximation of Fermi function}

\begin{figure}[hbtp]
\begin{center}
\includegraphics[width=70mm]{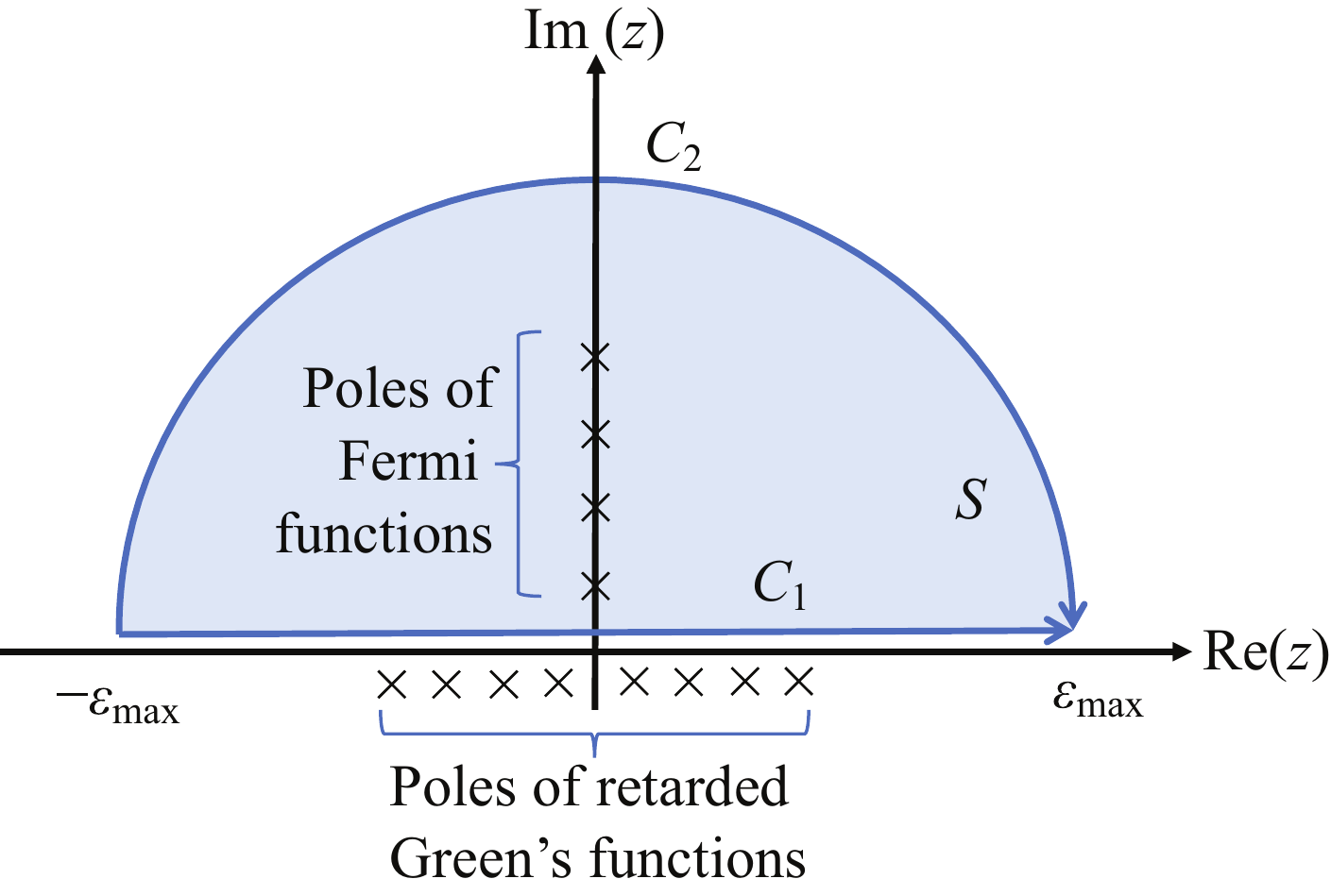}\\
\end{center}
\caption{Schematic of complex energy contour integration and poles of retarded Green's function and Fermi function. Note that the radius of the complex contour $C_{2}$ is taken large enough to avoid a steep change in the integrand.
\label{fig:CC}
}
\end{figure}

Because the integrand of Eq.~(\ref{eq:Jij_Green}) involves only the terms of the retarded Green's function, it is possible to reduce the computational cost by using the residue theorem instead of real axis integration.
The fundamentals of the complex contour integration of the retarded Green's function are illustrated in Fig.~\ref{fig:CC}.

To estimate the computational cost of energy integration, it is necessary to evaluate the steepness of the integrand. Because the Green's function, and thus the integrand, of Eq.~(\ref{eq:Jij_Green}) have poles just infinitesimally below the real axis (which is shown as $C_1$ in Fig.~\ref{fig:CC}), direct integration on the real axis would require very fine energy grid points.
Instead, the retarded Green's function changes gradually at the half-circle complex contour ($C_2$ in Fig.~\ref{fig:CC}), because the contour is far from all the poles of the retarded Green's functions.
From the residue theorem, the difference between integration on $C_1$ and on $C_2$ can be written as the summation of the residuals of the integrand:
\[
\int_{C_1}\mathcal{F}(z)dz-\int_{C_2}\mathcal{F}(z)dz
=2\pi i\sum_{z_p\in S}R_{p}[\mathcal{F}(z);z],
\]
where $\mathcal{F}(z)$ is the integrand,
$R_{p}[\mathcal{F}(z);z]$ is the residue of the integrand $\mathcal{F}(z)$ indexed by $p$ with the corresponding pole $z_{p}$, and $z_{p} \in S$ means that the pole $z_{p}$ is included in the closed area surrounded by $C_{1}$ and $C_{2}$ (pale blue area in Fig.~\ref{fig:CC}).
Given a sufficiently large radius $\varepsilon_{\mathrm{max}}$ of the contour $C_{2}$, it is possible to replace real axis integration with complex contour integration and summation of the residuals on the poles:
\begin{eqnarray}
J_{i\mathbf{0},j\mathbf{R}}
&=
  &\frac{1}{4\pi}\mathrm{Im}\left[\int_{C_1}\mathcal{F}(z)dz\right]
  \quad (\varepsilon_{\mathrm{max}}\rightarrow+\infty)
\nonumber \\
&=
 &\frac{1}{4\pi}\mathrm{Im}\left[\int_{C_2}\mathcal{F}(z)dz\right]
\nonumber \\
&&+\frac{1}{2}\mathrm{Re}\left\{\sum_{z_p\in S}R_{p}[\mathcal{F}(z);z]\right\},
\label{eq:Jij_contour}
\end{eqnarray}
with
\begin{eqnarray}
\mathcal{F}(z)
&=
 &f\left(\beta(z-\varepsilon_{\mathrm{F}})\right)
\nonumber \\
&&\times\sum_{\mu,\nu\in i}\sum_{\mu',\nu' \in j} 
  [\hat{P}_{i}]_{\nu\mu}
  G^{+}_{i\mu,j\nu'}(\downarrow,\varepsilon,\mathbf{R})
\nonumber \\
&&\quad\times[\hat{P}_{j}]_{\nu'\mu'}
  G^{+}_{j\mu',i\nu}(\uparrow,\varepsilon,-\mathbf{R}).
\end{eqnarray}

For the first term of Eq.~(\ref{eq:Jij_contour}), one can easily find that this term becomes zero at the limit $\varepsilon_{\mathrm{max}}\rightarrow +\infty$ (see Appendix \ref{append:infinite}).
By contrast, we have to consider all the poles of the Fermi function in the upper complex plane for the second term of Eq.~(\ref{eq:Jij_contour}). 
In this way, the Matsubara approximation, a standard approximation of the Fermi function, would result in slow convergence with respect to the number of poles. 
This is because the Matsubara approximation involves all the exact Fermi poles equally spaced on the imaginary axis, where the remaining part of the integrand decays as $z^{-2}$.  
Instead, 
we adopted the finite pole approximation of the Fermi function proposed by Ozaki~\cite{T_Ozaki_2007}.
That is, the approximated Fermi function can be written as in the summation of a finite number of fractions:
\begin{equation}
\tilde{f}_{N_{\mathrm{P}}}(z)
=\frac{1}{2}
+\sum_{p=1}^{N_{\mathrm{P}}}\frac{\tilde{R}_{p}}{z-\tilde{z}_{p}}
+\sum_{p=1}^{N_{\mathrm{P}}}\frac{\tilde{R}_{p}}{z+\tilde{z}_{p}}, 
\label{eq:Fermi_approx}
\end{equation}
where $z=\pm \tilde{z}_{p}$, and $\tilde{R}_{p}$ is the $p$-th positive/negative pole and the corresponding residue of the approximated Fermi function $\tilde{f}_{N_{\mathrm{P}}}(z)$.
This approximated Fermi function is derived from a hypergeometric function, and the poles and residues are calculated from the $N_{\mathrm{P}}$-dimensional generalized eigenvalue problem
\begin{equation}
\mathbf{Av}_{p}=\lambda_{p} \mathbf{Bv}_{p},
\quad (p=1,2,\ldots,N_{\mathrm{P}}),
\end{equation}
where $N_{\mathrm{P}}$ represents a sufficiently large number for convergence of the approximated Fermi function, and the elements of the matrices $\mathbf{A}$ and $\mathbf{B}$ can be defined as
\begin{eqnarray}
A_{qq'}
&=
 &-\frac{1}{2}(\delta_{(q+1)q'}+\delta_{(q-1)q'}),
\\
B_{qq'}
&=
 &\delta_{qq'}(2q-1),
\end{eqnarray}
where $q,q'=1,2,\ldots,N_{\mathrm{P}}$.
Using the eigenvalue $\lambda_{p}$ and eigenvector $v_{p}$, $\tilde{z}_{p}$ and $\tilde{R}_{p}$ can be written as
\begin{equation}
\tilde{z}_{p}
=\frac{i}{\lambda_{p}}
,\quad
\tilde{R}_{p}
=\frac{1}{4}v_{pp}^2\tilde{z}_{p}^2.
\end{equation}

In summary, Eq. (\ref{eq:Jij_Green}) can be rewritten as
\begin{eqnarray}
J_{i\mathbf{0},j\mathbf{R}}
&=
 &\frac{1}{2}
  \sum_{p=1}^{N_{\mathrm{P}}}\tilde{R}_{p}\sum_{\mu,\nu\in i}\sum_{\mu',\nu' \in j}
\nonumber \\
&&\quad\mathrm{Re}\left\{
  [\hat{P}_{i}]_{\nu\mu}
  G^{+}_{i\mu,j\nu'}(\downarrow,\tilde{z}_{p},\mathbf{R})
  \right.
\nonumber \\
&&\qquad\quad \times \left.
  [\hat{P}_{j}]_{\nu'\mu'}
  G^{+}_{j\mu',i\nu}(\uparrow,\tilde{z}_{p},-\mathbf{R})
  \right\}.
\label{eq:Jij_indiv_res}
\end{eqnarray}

\subsection{Liechtenstein formula for periodic images}

In addition to the exchange coupling constant of the individual site representation in Eq.~(\ref{eq:Jij_indiv_res}), it is possible to consider the exchange coupling constant $J_{ij}$ between the periodic images of $i$ and $j$, which is shown in Fig.~\ref{fig:sch_cells}(b).
This can be derived by summing up $J_{i\mathbf{0},j\mathbf{R}}$ for all the considered cell indices $\mathbf{R}$. That is,
\begin{equation}
J_{ij}\equiv\sum_{R}J_{i\mathbf{0},j\mathbf{R}}.
\label{eq:Jij_periodic_def}
\end{equation}
It is then possible to reduce the integration variable to only $\mathbf{k}$ from Eq.~(\ref{eq:Jij_eig}), as follows:
\begin{eqnarray}
J_{ij}
&=
 &\frac{1}{4}\int d^3\left(\frac{ka}{2\pi}\right)
  \sum_{n,n'}
  \frac{-f_{n\uparrow}(\mathbf{k})+f_{n'\downarrow}(\mathbf{k})}{\varepsilon_{n\uparrow}(\mathbf{k})-\varepsilon_{n'\downarrow}(\mathbf{k})}
\nonumber \\
&&\ \times \sum_{\mu,\nu\in i}\sum_{\mu',\nu' \in j}
   C_{j\mu',n\uparrow}(\mathbf{k})C^{*}_{i\nu,n\uparrow}(\mathbf{k})
\nonumber \\
&&\quad \times
  [\hat{P}_{i}]_{\nu\mu}
  C_{i\mu,n'\downarrow}(\mathbf{k})C^{*}_{j\nu',n'\downarrow}(\mathbf{k})
  [\hat{P}_{j}]_{\nu'\mu'},
\label{eq:Jij_periodic}
\end{eqnarray}
which is derived in Appendix \ref{append:allcell}.
This expression is useful when the unit cell is very large, because the interaction between the distant images of $i$ and $j$ [thin dashed lines in Fig.~\ref{fig:sch_cells}(b)] can be negligibly small compared with the interaction of the nearest images of $i$ and $j$ [thick solid line in Fig.~\ref{fig:sch_cells}(b)].

Hereafter, we refer to calculations of $J_{i\mathbf{0},j\mathbf{R}}$ using Eq.~(\ref{eq:Jij_indiv_res}) as individual pair calculations and those of $J_{ij}$ according to Fig.~\ref{fig:sch_cells}(b) as periodic image calculations to distinguish them.

\section{Computational models and methods}

\begin{figure}[hbtp]
\begin{center}
\includegraphics[width=60mm]{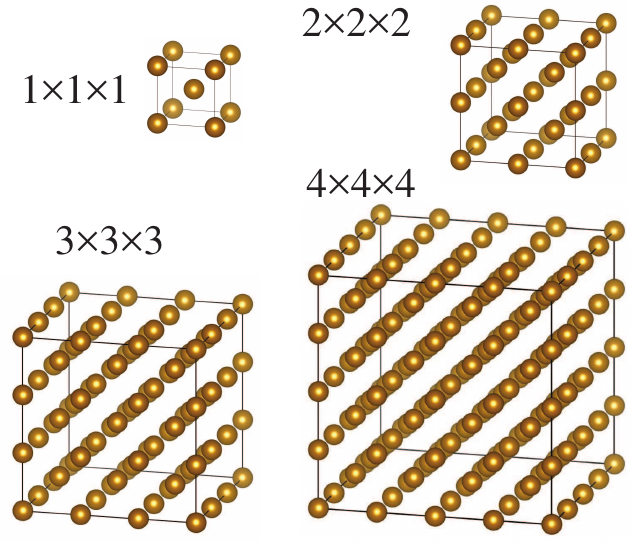}
\end{center}
\caption{Systems examined in this paper, which are bcc Fe structures with cell sizes ranging from $1\times 1 \times 1$ to $4 \times 4 \times 4$.
\label{fig:bccFe}
}
\end{figure}

Figure \ref{fig:bccFe} shows the five systems examined in this paper.
Each system consists of $N_{\mathrm{cell}}\times N_{\mathrm{cell}}\times N_{\mathrm{cell}}$ conventional unit cells of a bcc Fe crystal.

For first-principles calculation based on the localized basis set for the systems, we performed a density functional calculation using the OpenMX code \cite{T_Ozaki_2003}. Unless otherwise specified, we adopted the Perdew--Burke--Ernzerhof exchange-correlation functional \cite{GGA-PBE} within the generalized gradient approximation (GGA-PBE). For the pseudoatomic orbital basis sets, we adopted the $s2p2d2$ basis set for Fe, where the 3$p$, 3$d$ and 4$s$ states of Fe are treated explicitly as valence states.
The cutoff radius was set to 6.0 times the Bohr radius for Fe. 
We adopted the fully relativistic pseudopotentials generated by the Morrison--Bylander--Kleinman scheme \cite{MBK}.
We used an electronic temperature of 300 K, and the convergence criterion for the total energy was chosen as $1.0\times10^{-6}$ Ha.

The numbers of real and reciprocal space grids are determined according to the system size, as shown in Table \ref{tab:Nr_Nk}.
We set the numbers of real and reciprocal space grids proportional to the system size and the inverse of the system size, respectively, so as to obtain strict quantitative consistency between the calculation results for various system sizes.

\begin{table}[hbtp]
\begin{center}
\begin{tabular}{|c||c|c|c|c|}
\hline
System&$N_{r}\times N_{r}\times N_{r}$&$N_{k}\times N_{k}\times N_{k}$&$N_{\mathrm{atom}}$&$a$ [\AA]\\
\hline
$1\times 1 \times 1$ & $32 \times 32 \times 32$ & $24\times 24 \times 24$ & \ \,\ \,2 & \ \,2.866\\
$2\times 2 \times 2$ & $64 \times 64 \times 64$ & $12\times 12 \times 12$ & \ \,16 & \ \,5.732\\
$3\times 3 \times 3$ & $96 \times 96 \times 96$ & $8\times 8 \times 8$ & \ \,54 & \ \,8.598\\
$4\times 4 \times 4$ & $128 \times 128 \times 128$ & $6\times 6 \times 6$ & 128 & 11.464\\
\hline
\end{tabular}
\end{center}
\caption{
Numbers of real and reciprocal space grids, $N_{r}\times N_{r}\times N_{r}$ and $N_{k}\times N_{k}\times N_{k}$, for SCF calculations, together with the number of atoms and length of the unit cell vector for each system.
\label{tab:Nr_Nk}
}
\end{table}

\section{Results}

\subsection{Dependence of coupling constants on the computational parameters $N_{\mathrm{P}}$ and $N_{k}$}

After a self-consistent field (SCF) calculation is performed for each system, it is possible to obtain the Hamiltonian and overlap matrices corresponding to the converged electron density as an output of first-principles calculation.
The eigenvalues $\varepsilon_n(\mathbf{k})$ and vectors $\mathbf{C}(\mathbf{k})$ needed for calculations using Eqs.~(\ref{eq:g_up_2}) and (\ref{eq:g_down_2}) are obtained by solving the generalized eigenvalue problem represented by the Hamiltonian and overlap matrices.
It is then possible to calculate the exchange coupling constants using Eq.~(\ref{eq:Jij_indiv_res}).
Here, the computational accuracy of the $J_{ij}$ calculation is determined mainly by the number of $k$-points, which is $N_{k}\times N_{k}\times N_{k}$ for calculations using Eqs.~(\ref{eq:g_up_2}) and (\ref{eq:g_down_2}), and the number of poles $N_{P}$ in Eq.~(\ref{eq:Jij_indiv_res}).
To test the efficiency of the derived formalism in this paper, we examined the exchange coupling constant $J_{\mathrm{1NN}}$ between the first-nearest-neighbor sites in bcc Fe for various $N_{k}$ and $N_{\mathrm{P}}$ values. Note that the number of $k$-points in $J_{ij}$ is not necessarily the same as that in the SCF calculations. We employed a fixed value of $N_{k}$ for the SCF calculations, as shown in Table \ref{tab:Nr_Nk}, and examined different values of $N_{k}$ for the $J_{ij}$ calculations.

\begin{figure}[hbtp]
\begin{center}
\includegraphics[width=75mm]{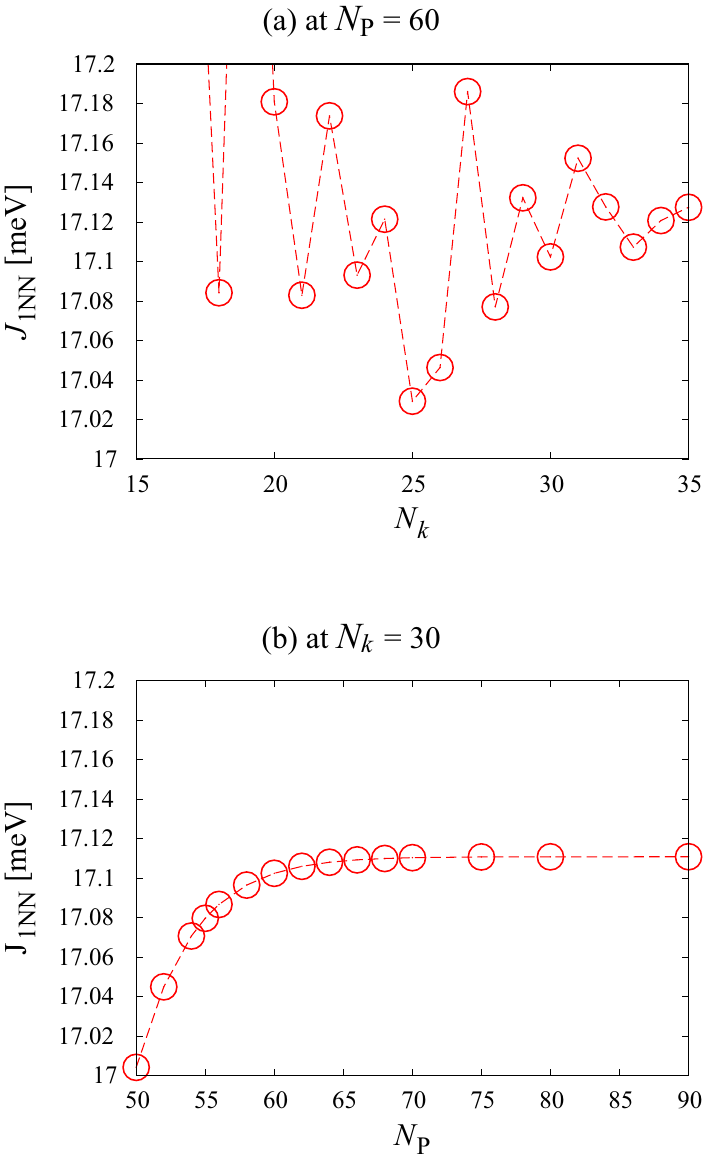}\\
\end{center}
\caption{
Exchange coupling constant $J_{\mathrm{1NN}}$ of first-nearest-neighbor pairs in the $1\times 1\times 1$ system (a) as a function of $N_{k}$ with a fixed $N_{\mathrm{P}}$ of 40 and (b) as a function of $N_{P}$ with a fixed $N_{k}$ of 30.
\label{fig:Nk_Np_J}
}
\end{figure}

Figure \ref{fig:Nk_Np_J} shows the calculated $J_{\mathrm{1NN}}$ values as functions of $N_{k}$ and $N_{\mathrm{P}}$ for the $1\times 1\times 1$ system. As shown in Fig.~\ref{fig:Nk_Np_J}(a), the $J_{\mathrm{1NN}}$ value oscillates within a range of approximately 0.1 meV even for a large $N_{k}$.
By contrast, $J_{\mathrm{1NN}}$ increases monotonically and converges within a range of 0.05 meV at around $N_{\mathrm{P}}=60$, as shown in Fig.~\ref{fig:Nk_Np_J}(b).
These features indicate that the computational cost of $J_{ij}$ calculation scales better for the number of poles $N_{k}$ than for the number of $k$-points, $N_{k} \times N_{k}\times N_{k}$.

This result must also be compared with the computational results obtained by adopting the Matsubara poles \cite{Y_O_Kvashnin_2015}.
As we discussed in Sec.~\ref{sec:Liech}, the Matsubara approximation would result in slow convergence on the order of $N_{\mathrm{P}}$, and the number of necessary Matsubara poles needed to obtain an accuracy of $10^{-5}$ Ry (0.136 meV) is 1024 at 300 K.
By contrast, Fig.~\ref{fig:Nk_Np_J}(b) shows that the calculated $J_{\mathrm{1NN}}$ value already converges to within 0.05 meV at $N_{\mathrm{P}}$. Thus, computation can be speeded up by a few ten times by straightforward adoption of the finite pole approximation of the Fermi function.

\subsection{Optimal computational parameters for $J_{ij}$ calculation}

To test the efficiency of our formalism for larger systems, we explore the optimal values of the computational parameters $N_{k}$ and $N_{\mathrm{P}}$ for various system sizes that give appropriate exchange coupling constants within an acceptable error tolerance.

For this purpose, $J_{\mathrm{1NN}}$ was again examined, and defined the acceptable error from the converged value as 0.05 meV for $J_{\mathrm{1NN}}$.

\begin{table}[hbtp]
\begin{center}
\begin{tabular}{|c||c|c|c|c|}
\hline
System & $N_{\mathrm{P}}^{\mathrm{opt}}$ & $N_{k}^{\mathrm{opt}}$ 
& $t_{\mathrm{eig}}$ [s] & $t_{\mathrm{indiv}}$ [s]\\
\hline
$1\times 1\times 1$ & 60 & 29 & \ \,\ \,1.5 & 1.9\\
$2\times 2\times 2$ & 60 & 15 & \ \,30.1 & 1.9\\
$3\times 3\times 3$ & 60 & \ \,8 & 193.2 & 1.3\\
$4\times 4\times 4$ & 60 & \ \,5 & 906.2 & 1.1\\
\hline
\end{tabular}
\end{center}
\caption{
Optimized values of computational parameters $N_{\mathrm{P}}$ and $N_{k}$ with corresponding computational times $t_{\mathrm{eig}}$ for eigenvalue calculation and $t_{\mathrm{indiv}}$ for $J_{ij}$ calculation based on Eq.~(\ref{eq:Jij_indiv_res}), and numbers of nodes and cores assigned in each calculation. All calculations are performed with MPI parallelization on 24 cores on 1 node of Intel(R) Xeon(R) E5-2680 v3 processors.
\label{tab:indiv}
}
\end{table}

Table \ref{tab:indiv} shows the set of smallest values $N^{\mathrm{opt}}_{\mathrm{P}}$ and $N^{\mathrm{opt}}_{k}$ that give the appropriate $J_{\mathrm{1NN}}$ with an error tolerance of 0.05 meV for each system.
We also show the computational times $t_{\mathrm{eig}}$ for eigenvalue calculation and the $t_{\mathrm{indiv}}$ value for the Liechtenstein calculation based on Eq.~(\ref{eq:Jij_indiv_res}), together with the numbers of nodes and cores. 
All calculations are performed by Intel(R) Xeon(R) E5-2680 v3 processors, and we parallelize only the calculations of the $k$-points using MPI and not the energy integrations. We adopt BLAS routines for the matrix multiplications and general eigenvalue problems.

Table \ref{tab:indiv} shows a few remarkable features.
The optimal value of $N_{\mathrm{P}}^{\mathrm{opt}}$ is almost independent of the system size, whereas the optimal value of $N^{\mathrm{opt}}_{k}$ decreases with increasing system size, and is roughly in inverse proportion to the system size.

It is also possible to compare the computational times $t_{\mathrm{eig}}$ required for the eigenvalue calculation with $t_{\mathrm{indiv}}$ for various system sizes.
When the system is small, $t_{\mathrm{eig}}$ is comparable $t_{\mathrm{indiv}}$, whereas $t_{\mathrm{eig}}$ increases quickly with increasing system size.
In contrast, $t_{\mathrm{indiv}}$ remains to the very small values even for large systems.
This means that it is possible to calculate $J_{ij}$ with small calculation costs even for large systems when the full set of eigenvalues and vectors for the necessary $k$ points is given.

\subsection{Comparison of $J_{ij}$ calculations for individual sites and periodic images}

Next, we compared the $J_{\mathrm{1NN}}$ values determined by Eq.~(\ref{eq:Jij_indiv_res}), which we call $J_{\mathrm{1NN}}$ for an individual pair, $J_{\mathrm{1NN}}^{\mathrm{indiv}}$, and those determined by Eq.~(\ref{eq:Jij_periodic}), which we call $J_{\mathrm{1NN}}$ for periodic images, $J_{\mathrm{1NN}}^{\mathrm{periodic}}$.
For quantitative comparison, we first defined the absolute difference $\Delta J_{\mathrm{1NN}}$ as
\begin{equation}
\Delta J_{\mathrm{1NN}}\equiv
\left|
  J_{\mathrm{1NN}}^{\mathrm{indiv}} - J_{\mathrm{1NN}}^{\mathrm{periodic}}
\right|.
\end{equation}

\begin{table}[hbtp]
\begin{center}
\begin{tabular}{|c||c|c|c|c|c|}
\hline
System& $\Delta J_{\mathrm{1NN}}$  [meV] & $t_{\mathrm{periodic}}$ [s]& $t_{\mathrm{indiv}}$ [s]\\
\hline
$1\times 1\times 1$ & 117.95 & 0.2 & 1.9 \\
$2\times 2\times 2$ & \ \,\ \,0.30 & 1.5 & 1.9 \\
$3\times 3\times 3$ & \ \,\ \,0.41 & 4.6 & 1.3 \\
$4\times 4\times 4$ & \ \,\ \,0.53 & 8.0 & 1.1 \\
\hline
\end{tabular}
\end{center}
\caption{
Difference $\Delta J_{\mathrm{1NN}}$ between the $J_{\mathrm{1NN}}$ values for periodic images and those for individual pairs in systems of various sizes, together with the computational times $t_{\mathrm{periodic}}$ and $t_{\mathrm{indiv}}$ of the periodic image and individual pair calculations, respectively.
For each system, the numbers of poles and $k$-points are the same as those for the individual pair calculations shown in Table \ref{tab:indiv}.
\label{tab:periodic}
}
\end{table}

Table \ref{tab:periodic} shows $\Delta J_{\mathrm{1NN}}$ for various system sizes, together with the computational times of the periodic image calculations, $t_{\mathrm{periodic}}$.
For consistency, $N_{k}$ and the computational conditions are set to the same values as those used in the individual pair calculations summarized in Table \ref{tab:indiv}.
We can see in Table \ref{tab:periodic} that $\Delta J_{\mathrm{1NN}}$ is much larger than $J_{\mathrm{1NN}}^{\mathrm{indiv}}$ itself in the $1\times 1\times 1$ system.
This is because in the periodic image calculation, all the interactions of neighboring sites in the $1\times 1\times 1$ system are counted, and the calculated value is approximately 8 times larger than the realistic value.
The difference $\Delta J_{\mathrm{1NN}}$ is much smaller for larger systems, whereas it exceeds the error tolerance of 0.05 meV.
Since the difference of 0.1 meV in $\Delta J_{\mathrm{1NN}}$ roughly corresponds to the deviation of 6 K in the Curie temperature, the differences of 0.3--0.5 meV seen in Tab.~\ref{tab:periodic} would result in errors of 18--31 K in the Curie temperature.
It is thus possible to conclude that the approximation of exchange coupling between individual sites to that of periodic images gives no advantage in terms of computational cost.



It is also necessary to note that the computational times $t_{\mathrm{eig}}$ for large systems are much larger than $t_{\mathrm{indiv}}$ and $t_{\mathrm{periodic}}$ for large systems. It is thus necessary to store certain sets of eigenvalues and vectors to speed up the computation of the coupling constants for number of pairs.

\subsection{Distance dependence of $J_{ij}$ and comparison with plane wave calculation}

Another important problem related to $J_{ij}$ calculation is that, we do not know what the most reliable $J_{ij}$ is.
Particularly, it is necessary to clarify whether $J_{ij}$ calculation derived in this study agrees quantitatively with those in other formulations. For this purpose, we calculated the distance dependence of $J_{ij}$, and compared it with the Liechtenstein calculation result obtained from Akai-KKR.

\begin{figure}[hbtp]
\begin{center}
\includegraphics[width=75mm]{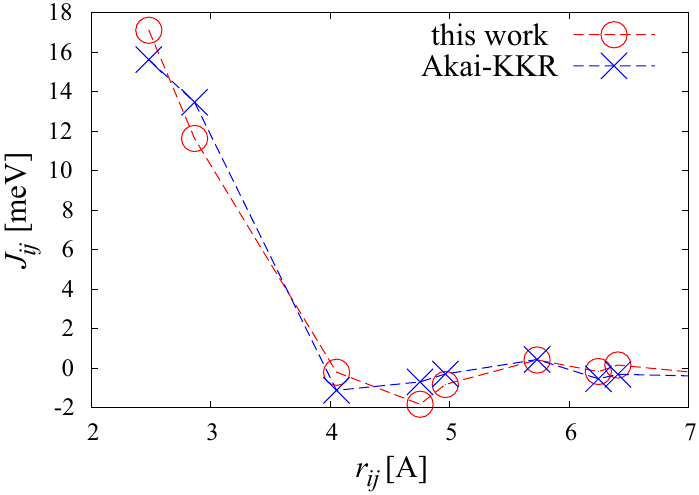}\\
\end{center}
\caption{
Exchange coupling constant as a function of atomic distance in a bcc Fe crystal. In this calculation, we adopted $N_{\mathrm{P}}=60$ and $N_{k}=30$ for the $1\times 1\times 1$ system.
\label{fig:r_vs_J}
}
\end{figure}

Figure \ref{fig:r_vs_J} shows $J_{ij}$ as a function of atomic distance $r_{ij}$ in a bcc-Fe crystal.
The red symbols represents $J_{ij}$ based on our formalism, and the blue symbols represents $J_{ij}$ from Akai-KKR.
Here, we adopted $N_{k}=30$ and $N_{\mathrm{P}}=60$ for $1\times 1\times 1$ system for the formalism in this study, and $N_{k}=20$, non-relativistic muffin-tin potential and GGA-PBE approximation in Akai-KKR calculation.
It is possible to see in Fig.~\ref{fig:r_vs_J} that the profiles of two results agree well with each other.

Moreover, it is possible to derive the Curie temperature from the calculated results. We calculated the Curie temperature $T_{\mathrm{C}}$ of bcc-Fe based on mean field approximation using computational results of $J_{i\mathbf{0},j\mathbf{R}}$ and $J_{ij}$ of system $1\times 1\times 1$ with Eqs.~(\ref{eq:Jij_indiv_res}) and (\ref{eq:Jij_periodic}), as the maximum eigenvalue of the matrix below:
\[
\frac{2}{3 k_{\mathrm{B}}}\left(\begin{array}{cc}
J_{11}-J_{1\mathbf{0},1\mathbf{0}}&J_{12}\\
J_{21}&J_{22}-J_{2\mathbf{0},2\mathbf{0}}
\end{array}\right).
\]
(For details, see Appendix \ref{sec:Curie}.)
As a result, the Curie temperature of bcc Fe is calculated as 1322 K. Considering the well known fact that the mean field approximation overestimates the Curie temperature of bcc Fe by $\sim$ 40\%, it is possible to say that our computational result is in a correspondence with the experimental value of 1043 K.

\section{Summary}

In this paper, we derived an explicit form of the Liechtenstein formula within the localized orbital basis representation and developed a computational code for the output of the first-principles calculation code OpenMX.
In the derivation, we adopted the finite pole approximation of the Fermi function, which simplified and speeded up the energy integrations.
To test the efficiency and computational speed of our implementation, we calculated the exchange coupling constant $J_{\mathrm{1NN}}$ of the first-nearest-neighbor pairs in bcc Fe crystals with various system sizes.
Using a new formalism based on the finite pole approximation, we were able to calculate $J_{\mathrm{1NN}}$ with an error tolerance of 0.05 meV with small computational times even for large systems.
To obtain an efficient formalism for calculating $J_{ij}$ for large systems, we compared the values of $J_{\mathrm{1NN}}$ based on the finite pole approximation and those of periodic image calculations for various system sizes.
It is shown that the approximation to periodic images gains almost no computational speedup, while the it gives the deviation of $J_{ij}$ at about 0.5 meV.
We also calculated the dependence of $J_{ij}$ on the atomic distance $r_{ij}$, and compared it with that obtained by Akai-KKR calculation.
It is shown that the two profiles agree well with each other, indicating the transferability of computational results derived by the two different formalisms.

\begin{acknowledgments}
The authors acknowledge Hisazumi Akai, Sonju Kou and Shotaro Doi for fruitful discussions and valuable comments.
This work was supported in part by MEXT, Japan, as a social and scientific priority issue CDMSI to be tackled by using the post-K computer, and the Elements Strategy Initiative Project under the auspices of MEXT, as well as KAKENHI Grant No.~17K04978. 
Some of the calculations were performed using the supercomputers at ISSP, The University of Tokyo, and TSUBAME, Tokyo Institute of Technology, as well as the K computer, RIKEN (Project Nos.~H30-Cb-0009, H31-Ca-0025, hp170269, hp180206, and hp190169).
One of the authors (MM)'s work in ISSP, Univ.~of Tokyo is supported by Toyota Motor Corporation.
\end{acknowledgments}

\appendix

\section{Detailed derivation of non-orthogonal Liechtenstein formula}\label{append:no_Liech}

In this section, we present a detailed derivation of Eqs.~(\ref{eq:Jij_eig}) and (\ref{eq:Jij_Green}) from the general form in Eq.~(\ref{eq:Jij_indiv}) by applying the completeness relationships for the Bloch functions $|\mathbf{k},n,\sigma\rangle$,
\begin{equation}
\sum_{\mathbf{k},n,\sigma}|\mathbf{k},n,\sigma\rangle\langle \mathbf{k},n,\sigma|=1, 
\label{eq:comp_Bl}
\end{equation}
and the completeness relationship for non-orthogonal basis sets,
\begin{equation}
\sum_{i,\mu,\mathbf{R}}\sum_{j,\nu,\mathbf{R'}}
|i,\mu,\mathbf{R}\rangle
[\mathbf{S}^{-1}_{\mathrm{RS}}]_{i\mu\mathbf{R},j\nu\mathbf{R}'}
\langle j,\nu,\mathbf{R}'|=1,
\label{eq:comp_rs}
\end{equation}
where $\mathbf{S}^{-1}_{\mathrm{RS}}$ is the overlap matrix of the real space basis.

Applying Eq. (\ref{eq:comp_Bl}) to Eq.~(\ref{eq:Jij_indiv}), we get
\begin{eqnarray}
&&J_{i\mathbf{0},j\mathbf{R}}
\nonumber \\
&&=
  \frac{1}{4\pi}
  \sum_{\mathbf{k},\mathbf{k}',\mathbf{k}'',\mathbf{k}'''}
  \sum_{n,n',n'',n'''}
  \sum_{\sigma,\sigma',\sigma'',\sigma'''}  
  \int d\varepsilon\,f\left(\beta(\varepsilon-\varepsilon_{\mathrm{F}})\right)
\nonumber \\
&&\quad\times\mathrm{ImTr}\left[
    \langle \mathbf{k}''',n''',\sigma'''|\hat{G}_{0\uparrow}^{+}(\varepsilon)|\mathbf{k},n,\sigma\rangle
    \right.
\nonumber \\
&&\qquad\qquad\,\times
  \langle \mathbf{k},n,\sigma|\hat{P}_{i,\mathbf{0}}|\mathbf{k}',n'\sigma'\rangle
\nonumber \\
&&\qquad\qquad\,\times
  \langle \mathbf{k}',n',\sigma'|\hat{G}_{0\downarrow}^{+}(\varepsilon)|\mathbf{k}'',n'',\sigma''\rangle
 \nonumber \\
&&\qquad\qquad\left.\times
  \langle\mathbf{k}'',n'',\sigma''|\hat{P}_{j,\mathbf{R}}|\mathbf{k}''',n''',\sigma'''\rangle
  \right].
\end{eqnarray}
Although the equation above has four sets of $(\mathbf{k},n)$, it is possible to drop two of them because the Green's function becomes diagonal in the eigenfunction representation, that is,
\begin{eqnarray}
&&\langle\mathbf{k}''',n''',\sigma'''|\hat{G}_{0\uparrow}^{+}(\varepsilon)|\mathbf{k},n,\sigma\rangle
\nonumber \\
&&=\frac{\delta_{n'''n}\delta_{\mathbf{k}'''\mathbf{k}}\delta_{\sigma'''\uparrow}\delta_{\sigma\uparrow}}
  {\varepsilon+i\eta-\varepsilon_{n\uparrow}(\mathbf{k})},
\\
&&\langle\mathbf{k}',n',\sigma|\hat{G}_{0\downarrow}^{+}(\varepsilon)|\mathbf{k}'',n'',\sigma''\rangle
\nonumber \\
&&=\frac{\delta_{n'n''}\delta_{\mathbf{k}'\mathbf{k}''}\delta_{\sigma'\downarrow}\delta_{\sigma''\downarrow}}
  {\varepsilon+i\eta-\varepsilon_{n'\downarrow}(\mathbf{k}')}.
\end{eqnarray}

The remaining task is to express the eigenfunction representations of the potential difference operators $\hat{P}_{i,\mathbf{0}}$ and $\hat{P}_{j,\mathbf{R}}$.
This can be done by applying Eq.~(\ref{eq:comp_rs}) as follows:
\begin{eqnarray}
&&\langle \mathbf{k},n,\uparrow|\hat{P}_{i,\mathbf{0}}|\mathbf{k}',n',\downarrow\rangle
\nonumber \\
&&=\sum_{i^{(1)},\mu^{(1)},\mathbf{R}^{(1)}}
  \sum_{j^{(1)},\nu^{(1)},\mathbf{R'}^{(1)}}
  \sum_{i^{(2)},\mu^{(2)},\mathbf{R}^{(2)}}
  \sum_{j^{(2)},\nu^{(2)},\mathbf{R'}^{(2)}}
\nonumber \\
&&\qquad
  \langle \mathbf{k},n,\uparrow|i^{(1)},\mu^{(1)},\mathbf{R}^{(1)}\rangle
\nonumber \\
&&\qquad\times
  [\mathbf{S}^{-1}_{\mathrm{RS}}]_{i^{(1)}\mu^{(1)}\mathbf{R}^{(1)},j^{(1)}\nu^{(1)}\mathbf{R}'^{(1)}}
\nonumber \\
&&\qquad\times
\langle j^{(1)},\nu^{(1)},\mathbf{R}'^{(1)}|\hat{P}_{i,\mathbf{0}}|i^{(2)},\mu^{(2)},\mathbf{R}^{(2)}\rangle
\nonumber \\
&&\qquad\times
  [\mathbf{S}^{-1}_{\mathrm{RS}}]_{i^{(2)}\mu^{(2)}\mathbf{R}^{(2)},j^{(2)}\nu^{(2)}\mathbf{R}'^{(2)}}
\nonumber \\
&&\qquad\times
  \langle j^{(2)},\nu^{(2)},\mathbf{R}'^{(2)}|\mathbf{k}',n',\downarrow\rangle.
\label{eq:Pi0_1}
\end{eqnarray}
Here we have to be careful that the localized basis representation of the potential difference operator, 
$\langle j^{(1)},\nu^{(1)},\mathbf{R}'^{(1)}|\hat{P}_{i,\mathbf{0}}|i^{(2)},\mu^{(2)},\mathbf{R}^{(2)}\rangle$,
becomes nonzero only at
$j^{(1)}=i^{(2)}=i$ and $\mathbf{R}'^{(1)}=\mathbf{R}^{(2)}=\mathbf{0}$.
Equation~({\ref{eq:Pi0_1}) reduces to
\begin{eqnarray}
&&\langle \mathbf{k},n,\uparrow|\hat{P}_{i,\mathbf{0}}|\mathbf{k}',n',\downarrow\rangle
\nonumber \\
&&=\sum_{i^{(1)},\mu^{(1)},\mathbf{R}^{(1)}}
  \sum_{\nu,\mu\in i}
  \sum_{j^{(2)},\nu^{(2)},\mathbf{R'}^{(2)}}
\nonumber \\
&&\qquad
  \langle \mathbf{k},n,\uparrow|i^{(1)},\mu^{(1)},\mathbf{R}^{(1)}\rangle
\nonumber \\
&&\qquad\times
  [\mathbf{S}^{-1}_{\mathrm{RS}}]_{i^{(1)}\mu^{(1)}\mathbf{R}^{(1)},i\nu\mathbf{0}}
  \langle i,\nu,\mathbf{0}|\hat{P}_{i,\mathbf{0}}|i,\mu,\mathbf{0}\rangle
\nonumber \\
&&\qquad\times
  [\mathbf{S}^{-1}_{\mathrm{RS}}]_{i\mu\mathbf{0},j^{(2)}\nu^{(2)}\mathbf{R}'^{(2)}}
\nonumber \\
&&\qquad\times
  \langle j^{(2)},\nu^{(2)},\mathbf{R}'^{(2)}|\mathbf{k}',n',\downarrow\rangle.
\label{eq:Pi0_2}
\end{eqnarray}
One can then replace the Bloch functions with the localized orbital representation in Eq.~(\ref{eq:PAO}) to obtain
\begin{eqnarray}
&&\langle \mathbf{k},n,\uparrow |i^{(1)},\mu^{(1)},\mathbf{R}^{(1)}\rangle
\nonumber \\
&&=\frac{1}{\sqrt{N}}\sum_{j',\nu',\mathbf{R}'}e^{-i\mathbf{k}\cdot\mathbf{R}'}
  C^{*}_{j'\nu',n\uparrow}(\mathbf{k})
\nonumber \\
&&\qquad\times
  [\mathbf{S}_{\mathrm{RS}}]_{j'\nu'\mathbf{R}',i^{(1)},\mu^{(1)},\mathbf{R}^{(1)}}
\label{eq:Pi0_3} \\
&&\langle j^{(2)},\nu^{(2)},\mathbf{R}^{(2)}|\mathbf{k}',n',\downarrow\rangle
\nonumber \\
&&=\frac{1}{\sqrt{N}}\sum_{i'',\mu'',\mathbf{R}''}
  [\mathbf{S}_{\mathrm{RS}}]_{j^{(2)}\nu^{(2)}\mathbf{R}^{(2)},i''\mu''\mathbf{R}''}
\nonumber \\
&&\qquad\times
  e^{i\mathbf{k}'\cdot\mathbf{R}''}C_{i''\mu'',n'\downarrow}(\mathbf{k}').
\label{eq:Pi0_4}
\end{eqnarray}
Combining Eqs.~(\ref{eq:Pi0_2}), (\ref{eq:Pi0_3}), and (\ref{eq:Pi0_4}), we obtain
\begin{eqnarray}
&&\langle \mathbf{k},n,\uparrow|\hat{P}_{i,\mathbf{0}}|\mathbf{k}',n',\downarrow\rangle
\nonumber \\
&&=\frac{1}{N}\sum_{\nu,\mu\in i}
  C^{*}_{i\nu,n\uparrow}(\mathbf{k})
  [\hat{P}_{i}]_{\nu\mu}
  C_{i\mu,n'\downarrow}(\mathbf{k}'),
\end{eqnarray}
with $[\hat{P}_{i}]_{\nu\mu}\equiv\langle i,\nu,\mathbf{0}|\hat{P}_{i,\mathbf{0}}|i,\mu,\mathbf{0}\rangle$. Similarly,
\begin{eqnarray}
&&\langle \mathbf{k}',n',\downarrow|\hat{P}_{j,\mathbf{R}}|\mathbf{k},n,\uparrow\rangle
\nonumber \\
&&=\frac{1}{N}\sum_{\nu',\mu'\in j}
  C^{*}_{j\nu',n'\downarrow}(\mathbf{k})e^{-i\mathbf{k}'\cdot\mathbf{R}}
\nonumber \\
&&\quad \times
  [\hat{P}_{j}]_{\nu'\mu'}
  e^{i\mathbf{k}\cdot\mathbf{R}}C_{i\mu',n\uparrow}(\mathbf{k}),
\end{eqnarray}
with $[\hat{P}_{j}]_{\nu'\mu'}\equiv\langle j,\nu',\mathbf{R}|\hat{P}_{j,\mathbf{R}}j,\mu',\mathbf{R}\rangle$.
The resulting equation can be written in explicit form as
\begin{eqnarray}
&&J_{i\mathbf{0},j\mathbf{R}}
\nonumber \\
&&=\frac{1}{4\pi N^2}\sum_{\mathbf{k},\mathbf{k}'}
  \sum_{n,n'}\sum_{\nu,\mu\in i}\sum_{\nu',\mu'\in j}
  \int d\varepsilon\,f\left(\beta(\varepsilon-\varepsilon_{\mathrm{F}})\right)
\nonumber \\
&&\times\mathrm{ImTr}\left[
    \frac{
      C^{*}_{i\nu,n\uparrow}(\mathbf{k})
      [\hat{P}_{i}]_{\nu\mu}
      C_{i\mu,n'\downarrow}(\mathbf{k}')}
      {\varepsilon+i\eta-\varepsilon_{n\uparrow}(\mathbf{k})}
   \right.
\nonumber \\
&&\left.\times
  \frac{
    C^{*}_{i\nu',n'\downarrow}(\mathbf{k}')e^{-i\mathbf{k}'\cdot\mathbf{R}}
    [\hat{P}_{j}]_{\nu'\mu'}
    e^{i\mathbf{k}\cdot\mathbf{R}}C_{i\mu',n\uparrow}(\mathbf{k})
  }{\varepsilon+i\eta-\varepsilon_{n'\downarrow}(\mathbf{k}')}
  \right].
\end{eqnarray}
Because we treat periodic systems, the matrix representations of $\hat{P}_{i,\mathbf{0}}$ and $\hat{P}_{j,\mathbf{R}}$ by localized basis sets do not depend on the cell indices $\mathbf{0}$ and $\mathbf{R}$.
Finally, we obtain Eq.~(\ref{eq:Jij_indiv_loc}) by replacing the summations of $\mathbf{k}$ and $\mathbf{k}'$ with the corresponding integrals.

\section{Complex contour integration at the limit $\varepsilon_{\mathrm{max}}\rightarrow +\infty$}\label{append:infinite}

In this subsection, we consider the behavior of the integrand $\mathcal{F}(z)$ in Eq.~(\ref{eq:Jij_contour}) on the complex contour $C_{2}$ at the limit $\varepsilon_{\mathrm{max}}\rightarrow +\infty$.
From the conditions, we can straightforwardly derive $|z|\rightarrow +\infty$ on $C_{2}$, and the energy eigenvalues and poles become negligible compared with $|z|$:
\begin{equation}
z\in C_{2} \Rightarrow |z|\gg |\varepsilon_{n}|, |\tilde{z}_{p}|.
\label{eq:inf_1}
\end{equation}
Then it is possible to take the limit of the fractions in the Green's functions and Fermi functions as follows:
\begin{eqnarray}
\frac{1}{z+i\eta-\varepsilon_{n}(\mathbf{k})}
&=
 &\frac{1}{z}\cdot\frac{1}{1-\delta}
\nonumber \\
&=
 &\frac{1}{z}\left(
   1+\delta+\delta^2+\delta^2+\cdots
  \right),
\label{eq:inf_2}
\end{eqnarray}
\begin{equation}
\frac{1}{z-\tilde{z}_{p}}
=\frac{1}{z}\left(
   1+\delta'+\delta'^2+\delta'^2+\cdots
  \right),
\label{eq:inf_3}
\end{equation}
where
\[
\delta=\frac{\epsilon_{n}(\mathbf{k})-i\eta}{z},\quad
\delta'=\frac{z_{p}}{z}.
\]
It is straightforward to derive $|\delta|,|\delta'|\rightarrow 0$ from Eq.~(\ref{eq:inf_1}), and we can neglect the first- and higher-order terms of $\delta$ and $\delta'$ in Eqs.~(\ref{eq:inf_2}) and (\ref{eq:inf_3}).
It is then possible to find that
\begin{equation}
\mathcal{F}(z)
\propto
\left(
  \frac{1}{2}
  +2\sum_{p=1}^{N_{\mathrm{P}}}\frac{\tilde{R}_{p}}{z}
\right)\frac{1}{z^2}
\propto \frac{1}{z^2} 
\quad (|z|\rightarrow +\infty).
\end{equation}
Consequently, the integration of $\mathcal{F}(z)$ on $C_{2}$ yields a value of zero.

\section{Summation for periodic images}\label{append:allcell}
In this section, we reduce the double integration of $\mathbf{k}$ in Eq.~\ref{eq:Jij_eig} to a single integration by taking the summation of periodic images shown in Fig.~\ref{fig:sch_cells}(b).
From the definition in Eq.~(\ref{eq:Jij_periodic_def}),
\begin{eqnarray}
&&J_{ij}
\nonumber \\
&&=\frac{1}{4}\int d^3\left(\frac{ka}{2\pi}\right)\int d^3\left(\frac{ka}{2\pi}\right)'\left[
   \sum_{R}e^{i(\mathbf{k}-\mathbf{k}')\cdot\mathbf{R}}
  \right]
\nonumber \\
&&\quad\times\,\sum_{n,n'}
  \sum_{\mu,\nu\in i}\sum_{\mu',\nu' \in j}
  \frac{-f_{n\uparrow}(\mathbf{k})+f_{n'\downarrow}(\mathbf{k}')}{\varepsilon_{n\uparrow}(\mathbf{k})-\varepsilon_{n'\downarrow}(\mathbf{k}')}
\nonumber \\
&&\qquad \times\,\mathrm{Tr}\left[
   C_{j\mu',n\uparrow}(\mathbf{k})C^{*}_{i\nu,n\uparrow}(\mathbf{k})
  [\hat{P}_{i}]_{\nu\mu} \right.
\nonumber \\
&&\qquad\quad \left.\times
  C_{i\mu,n'\downarrow}(\mathbf{k}')C^{*}_{j\nu',n'\downarrow}(\mathbf{k}')
  [\hat{P}_{j}]_{\nu'\mu'}
  \right].
\end{eqnarray}
Considering the relationship
\begin{eqnarray*}
\sum_{R}e^{i(\mathbf{k}-\mathbf{k}')\cdot\mathbf{R}}
&=
 &\frac{1}{V}\int d^3 r
  \exp[i(\mathbf{k}-\mathbf{k}')\cdot \mathbf r]
\nonumber \\
&=
 &\frac{(2\pi)^3}{V}\delta^3(\mathbf{k}-\mathbf{k}'),
\end{eqnarray*}
we obtain Eq.~(\ref{eq:Jij_periodic}).

\section{Curie temperature of periodic systems}\label{sec:Curie}

Using the mean field approximation, Curie temperature of general periodic system can be obtained by finding the maximum value of $T$ which satisfies the following relationship:
\begin{eqnarray}
T\langle\vec{s}_{i}\rangle_{z}
&=
 &\frac{2}{3 k_{\mathrm{B}}}
  \sum_{j\mathbf{R}\ne i\mathbf{0}}J_{i\mathbf{0},j\mathbf{R}}\langle\vec{s}_{j}\rangle_{z}.
\end{eqnarray}
Using the definition in Eq. (\ref{eq:Jij_periodic}), the relationship can be rewritten as
\begin{eqnarray}
T\langle\vec{s}_{i}\rangle_{z}
&=
 &\frac{2}{3 k_{\mathrm{B}}}\sum_{j}\tilde{J}_{ij}\langle\vec{s}_{j}\rangle_{z}.
\\
\tilde{J}_{ij}
&\equiv
 &J_{ij}-J_{i\mathbf{0},j\mathbf{0}}\delta_{ij}.
\end{eqnarray}
That is, because $J_{i\mathbf{0},i\mathbf{0}}$ derived by Liechtenstein formula need not to become zero but doesn't have physical meaning, it is necessary to subtract its contribution from the periodic sum.

\end{document}